# A mechanism of carrier doping induced magnetic phase transitions in two-dimensional materials

Yan Lu[1], Haonan Wang[2], Li Wang[1], and Li Yang[2]

1. Department of Physics, Nanchang University, Nanchang, 330031, People's Republic of China

2. Department of Physics and Institute of Materials Science and Engineering, Washington University, St. Louis, MO, 63130, USA

## Abstract

Electrically tuning long-range magnetic orders has been realized in two-dimensional (2D) semiconductors via electrostatic doping. On the other hand, the observations are highly diverse: the transition can be realized by either electrons or holes or both depending on specific materials. Moreover, doped carriers seem to always favor the ferromagnetic (FM) ground state. The mechanism behind those diverse observations remains uncovered. Combining first-principles simulations, we analyze the spin superexchange paths of the correlated *d/f* orbitals around band edges and assign 2D magnetic semiconductors into three types by their projected density of states (PDOS). We find that each type of PDOS corresponds to a specific carrier-driven magnetic phase transition and the critical doping density and type of carriers can be quantitatively obtained by calculating the superexchange coupling strength. The model results are in good agreements with first-principles calculations and available measurements. After understanding the mechanism, we can design heterostructures to realize the FM to antiferromagnetic transition, which has not been realized before. This model is helpful to understand diverse measurements and expand the degrees of freedom to control long-range magnetic orders in 2D semiconductors.

## I. Introduction

Tuning magnetic properties by electrical methods is highly desirable due to the advantages of space confinement, easy manipulation, energy efficiency, and quick response [1]. Because of the unique atomic thickness, electric gating is particularly efficient for controlling electronic properties and free carriers in two-dimensional (2D) materials [2]. To date, numerous theoretical calculations and experiments have shown that doped free carriers not only modify magnetic transition temperature [3-5], hysteresis loop [3,6], anisotropy [4,7], but also are able to control the long-range magnetic orders [3,8,9]. On the other hand, the observations of carrier-driven magnetic transitions are diverse. For example, the interlayer antiferromagnetic (AFM) coupling of bilayer $CrI_3$ can be tuned to ferromagnetic (FM) coupling by electron doping, while the magnetic order is insensitive to hole doping [3,8]. On the contrary, it is predicted that the interlayer coupling can be tuned to be FM coupling by hole doping in $MnBi_2Te_4$ (MBT) through doping various nonmagnetic p-type elements [10], but electron doping cannot. Moreover, both electron and hole dopings have been predicted to be effective to tune the magnetic order of α-$RuCl_3$ [11], a candidate material for quantum spin liquid (QSL) [12-17].

These measurements and predictions raise the questions that how doped carriers change the magnetic coupling, and why sometimes doped electrons and holes induce symmetric while other times induce asymmetric effects in tuning the magnetic couplings. Numerous mechanisms can be applied to understanding these results [18,19]. For example, the free carrier induced itinerant magnetism is reasonable for understanding the observed AFM/FM transition but it falls to explain why only one type of carriers works in many materials [3,8]. Therefore, a broad picture is still lacking in this field although it is crucial for efforts to purposely design material magnetic properties by the efficient gating/doping.

In this work, we propose a model to understand the role of doped carriers in tuning the magnetic coupling of 2D magnetic materials. Based on the picture of spin superexchange, we show that there are three types of density of states (DOS) depending on the distribution of correlated *d* orbitals around the band edges. Consequently, this results in three types of doping tunable magnetic couplings in 2D magnetic semiconducting materials. Using first-principles density functional theory (DFT) calculations, we show that this model quantitatively agrees with the DFT-calculated changes of magnetic states in a wide range of 2D magnetic materials, such as bilayer $CrI_3$, bilayer $MnBi_2Te_4$, and monolayer α-$RuCl_3$. Finally, employing this mechanism, we propose to tune interlayer FM heterostructure to the AFM order by carrier doping, which has not been realized.

The article is organized in the following way. In section II, we present the computational details. In section III, we present the physical origins of tuning magnetic exchange coupling by carrier doping, which can be divided into three types according to their DOS. In Section IV, we verify these three types by DFT calculations of bilayer $CrI_3$, bilayer $MnBi_2Te_4$, and monolayer α-$RuCl_3$. In Section V, we expand this picture to predict the FM to AFM transition by carrier doping. The results are concluded in Section VI.

## II. Calculation Methods

Calculations are performed by using first-principles DFT, as implemented in *Vienna Ab initio Simulation Package* (VASP) [20,21]. Plane-wave basis set with a kinetic energy cutoff of 400 eV is adopted for geometry optimizations and self-consistent calculations. For geometry optimizations, all atoms are fully relaxed until the residual force per atom is less than 0.01 eV/Å. A vacuum distance larger than 12 Å is set between adjacent slabs to avoid spurious interactions. Spin-orbital coupling (SOC) is included in structural relation and total-energy calculations in this work.

Bilayer $CrI_3$ are calculated by applying dispersion-corrected PBEsol functionals [22]. On-site Coulomb interactions of *d* orbitals of Cr are adopted by Liechtenstein scheme [23] with U = 3.9 eV and J = 1.1 eV, as successfully used in previous calculations [24].

The first Brillouin zone is sampled with 9×9×1 k-point meshes. Monolayer RuCl$_3$, bilayer MBT, and TiBi$_2$Te$_4$/NiBi$_2$Te$_4$ heterostructure are calculated by applying PBE functional [25]. On site Coulomb interactions are adopted by the Dudarev scheme [26], with U = 2.0, 5.34, 3.0, and 4.0 eV for Ru, Mn, Ti, and Ni, respectively, which have been successful applied in previous calculations[11,27,28]. Van der Waals (vdW) interactions are adopted by the DFT-D3 method [29]. The first Brillouin zones are sampled with 15×9×1, 21×21×1, and 21×21×1 k-point meshes for monolayer RuCl$_3$, bilayer MnBi$_2$Te$_4$, and TiBi$_2$Te$_4$/NiBi$_2$Te$_4$ heterostructure, respectively.

### III.  Variation of the Superexchange coupling by Carrier Doping

There are many types of magnetic interactions, such as direct exchange, indirect superexchange, and indirect double exchange, etc [30]. For our studies 2D compounds, the direct exchange is not important, because the *d/f* orbitals of magnetic ions are strongly localized, and their nearest site are non-magnetic ions. The double exchanges can also be neglected since there are no mixed valence states for these materials. Thus, we adopt the local spin moments and consider the spin superexchange as the main factor in deciding the magnetic order. Spin superexchange is an indirect exchange interaction between non-neighboring magnetic ions, and it is mediated by intermediate states (typically nonmagnetic *p* orbitals) between magnetic ions [31-33]. The spin superexchange arise because electron may acquire additional kinetic energy advantage by hopping between the local magnetic ions [30].

Electron or hole doping will introduce partially occupied states, which open additional spin superexchange paths through partially occupied states at band edges. We begin our discussion by considering the spin superexchange between two partially occupied spin-polarized *d* orbitals. Fig. 1(a) shows an example of the additional superexchange interaction between two partially occupied *d* orbitals under carrier doping. Because of the requirement of spin conservation, spins can hop from one to another *d* orbitals with the same spin direction, resulting in a FM coupling. For the same reason, the spins cannot hop between *d* orbitals with opposite spin directions (the AFM coupling). As a result, the FM coupling is typically more energy favorable than the AFM coupling between two partially occupied *d* orbitals.

Based on the above discussion, the doping effect mainly affects the FM coupling. We can quantitatively estimate the favorite energy of the FM coupling by calculating the strength of additional FM superexchange paths. For the additional superexchange path shown in Fig. 1(a), the favorite FM strength can be written as [31-33]

$$\Delta J_S = \langle \Psi_d^L | H_{dp} | \Psi_p \rangle \langle \Psi_p | H_{pd} | \Psi_d^R \rangle, \quad (1)$$

$$= \rho_d \rho_p t_{dp}^2, \quad (2)$$

where $|\Psi_d^L\rangle$, $|\Psi_p\rangle$, and $|\Psi_d^R\rangle$ are the wavefunctions of the left *d*, intermedia *p*, and right *d* orbitals in Fig. 1(a), respectively. $H_{dp}$ and $H_{pd}$ are the hopping Hamiltonians at left and right part, respectively. $\rho_d$ and $\rho_p$ are the projected density of states (PDOS) of the *d* and *p* orbitals around the Fermi level, respectively. $t_{dp}$ is the hopping parameter between the *d* and *p* orbitals, which depends on their wavefunctions overlap and is usually independent on doping or band occupation. Therefore, Eq. (2) shows that the favorite energy is mainly decided by the product of PDOS ($\rho_d$ and $\rho_p$) around Fermi level.

Importantly, the electronic bands of *p* orbitals are usually dispersive, resulting in smoothly varied PDOS. On the contrary, those of correlated *d* orbitals are flatter, forming drastic change of PDOS. Moreover, usual doping only affects electronic states and occupation around the Fermi level or band edge of semiconductors. As a result, we conclude that those *d* orbitals around the band edge are crucial for the carrier-driven magnetic properties. Following this idea, we categorize 2D magnetic semiconductor materials into three types, I, II, and III, according to the distribution of *d* orbitals around band edges, as shown in Figs 1(b-d).

For the type-I PDOS shown in Fig. 1(b), there are *d* orbitals at the conduction edge but not at the valence edge. Therefore, electron doping can introduce partially occupied *d* orbitals and favor the FM coupling. Meanwhile, hole doping does not affect the magnetism because there is no *d*-orbital component around the valence band edge and the change of PDOS is smooth. The similar picture can be applied to understand the type-II PDOS shown in Fig. 1(c), in which there are *d* orbitals at the valence edge but not at the conduction edge. Thus, we expect only hole doping can introduce the partial occupied *d* orbitals and tune the superexchange strength. Finally, for the type III PDOS shown in Fig. 1(b), there are *d* orbitals at both the conduction and valence band edges. Therefore, both electron and hole doping can introduce partially occupied *d* orbitals at conduction and valence edges. As a result, both electron and hole doping may dramatically change the PDOS around the Fermi level and favor the FM couplings.

### IV.    Doping-induced AFM to FM of Magnetic Transition

To quantitatively demonstrate this picture, we will employ first-principles simulations to calculate the doping-induced change of magnetic couplings of 2D magnetic materials. For each type of PDOS, we choose the representative materials, including bilayer $CrI_3$, bilayer MBT, and monolayer α-$RuCl_3$,

Bilayer $CrI_3$: Few-layer $CrI_3$ is the first set of 2D magnetic materials achieved in experiments [34]. It may be used as possible magnetic information storage, due to their drastically enhanced tunneling magnetoresistance with layer thickness [35-38]. This material also attracts significant attentions owing to its possibility of tuning magnetic order by magnetic field [34], gating [3,8,39], pressure [40,41] and interlayer stacking [42-48]. The magnetic ground state of 2D $CrI_3$ is intralayer FM but interlayer AFM, as shown in the left part of Fig. 2(b). Particularly, it is observed that electron doping can tune the interlayer AFM to FM coupling at a doping density of $2.5 \times 10^{13}$ cm$^{-2}$, but interestingly, hole doping cannot [3,8]. However, the mechanism for this electron-doping induced magnetic phase transition has not been well explained.

In our DFT simulation, we adopt the monoclinic interlayer stacking phase because experiments show that mechanical exfoliation bilayer $CrI_3$ always stay at this high temperature phase [42-44], which is formed by a shift of one third of the in-plane lattice constant from the AA stacking style along the zigzag direction, as shown in Fig. 2(a). We find the magnetic ground state of this bilayer $CrI_3$ is intralayer FM but interlayer AFM. This agrees with measurements and previous works [45-48]. Both interlayer FM and AFM orders are schematically presented in Fig. 2(b). Our calculated total energy of the interlayer AFM order is about 0.24 meV lower than that of the interlayer FM coupling. This small energy difference between interlayer FM and AFM couplings also gives hope to tune the magnetic order via carrier doping.

The PDOS of bilayer FM $CrI_3$ is plotted in Fig. 2(c). We find that there are significant *d* orbitals at the conduction edge but nearly no *d* orbitals at the valence edge. Therefore, the PDOS of bilayer $CrI_3$ belongs to type I, and we expect electron doping may be effective to realize the FM interlayer coupling. The DFT calculation shown in Fig. 2(d) confirms this expectation. Doped electrons can change the magnetic states from interlayer AFM to FM states at a critical density of 0.01e/u.c. while doped holes have no effect.

We can further obtain the quantitative energy change by the model based on Eq. (2) with the input PDOS from first-principles calculations. Eq. (2) shows that the energy difference is $\Delta E = t_{dp}^2 \rho_d \rho_p - \Delta E_0$, in which $\Delta E_0$ is the energy difference of the undoped (intrinsic) structure and it is 0.24 meV for bilayer $CrI_3$. First-principles results in Fig. 2 (c) can provide the PDOS ($\rho_d$ and $\rho_p$) at the Fermi level, and they are plotted in Fig. 2 (e) according to the doping density. As expected, the PDOS of *p* orbitals are smooth while that of *d* orbitals shows an asymmetric behavior according to the doping type. Finally, the model result is plotted in Fig. 2 (d), and it is in an excellent agreement with first-principles results. The only fitting parameter is the hopping $t_{dp}^2$, which is 0.5 eV$^3$.

<u>Bilayer MnBi$_2$Te$_4$</u>: Layered MBT may displays rich topological properties from an axion insulator to a quantum anomalous Hall insulator [28,49-51]. To this end, interlayer FM coupling is highly desired, but unfortunately, DFT calculations and experiments showed that the interlayer is AFM coupling, which is about 0.43 meV lower than that of interlayer FM coupling [28], as shown in Fig. 3(a). This spurs significant efforts to find approaches to tune the interlayer couplings from AFM to FM in layered MBT [52-56]. In this section, we show that doping can effectively tune the interlayer magnetic order and it can be explained by our proposed mechanism.

The DFT-calculated PDOS of bilayer MBT with the FM interlayer coupling is shown in Fig. 3(b). There are $d$ orbitals at the valence band edge, but there is nearly no $d$-orbital component at the conduction band edge. Therefore, its PDOS belongs to type II. We expect that hole doping can effectively tune the magnetic order while electron doping does not. The first-principles DFT result is presented in Fig. 3(c). As expected, a critical hole doping density of 0.06 h/u.c. can change the magnetic states from interlayer AFM to FM state while doped electrons do not affect the magnetism.

We estimate this doping-induced total energy change by the model of Eq (2). For this purpose, the change of PDOS of $d$ orbitals of Mn ions, $\rho_d$, those of $p$ orbitals from Bi and Te ions, $\rho_p$, induced by doping are shown in Fig. 3(d). According to Eq. (2), we obtain the formula, $\Delta E = [t_{dp}^2 \rho_{Mn(d)}(\rho_{Bi(p)} + \rho_{Te(p)}) - \Delta E_0]$, in which $\Delta E_0$ is the energy difference of the undoped (intrinsic) structure and it is 0.43 meV for bilayer MBT. When we choose the fitting parameter $t_{dp}^2 = 7$ eV$^3$, Fig. 3(d) shows that the model excellently agrees with the DFT-calculated energy difference.

Monolayer α-RuCl$_3$: Bulk α-RuCl$_3$ is a layered material with layers stacked by weak van der Waals (vdW) interactions. It is regarded as a promising candidate to realize quantum spin liquid and topological Majorana modes [12-17]. The structure of each monolayer α-RuCl$_3$ is shown in Fig. 4(a). The DFT-calculated magnetic ground state of monolayer α-RuCl$_3$ is the so-called "zigzag antiferromagnetic" (zAFM) [11], whose spin configurations are shown in the left part of Fig. 4(a). A metastable state is the FM state, as shown in the right part of Fig. 4(a), whose energy are about 2.6 meV higher than that of zAFM in our DFT calculations, which is consistent with previous calculations [11,57-59]. The calculated PDOS of monolayer FM α-RuCl₃ is shown in Fig. 4(b), from which we can see that there are $d$ orbitals at both conduction and valence edges. Therefore, the PDOS of monolayer α-RuCl₃ belongs to type III, and our previous discussion expects that both doped electrons and holes will favor the FM state in this material.

Indeed, our DFT calculations shown that both electron and hole doping can tune the magnetic ground state from zAFM to FM, as shown by the orange and green curves shown in Fig. 4(c). The critical density is about 0.05 electrons per unit cell (~$3.02 \times 10^{13}$ cm⁻²) and 0.02 holes per unit cell (~$1.21 \times 10^{13}$ cm⁻²), respectively. Moreover, hole doping is more effective to change the magnetic coupling in monolayer α-RuCl₃. These first-principles results are consistent with previous works [11].

Similarly, we can obtain the change of superexchange strengths according to Eq. (2) under carrier doping. The PDOS of $\rho_d$ and $\rho_p$ vs the doping density is plotted in Fig. 4(d). Using the hopping parameter $t_{dp}^2 = 7.2$ eV³, the model is in a very good agreement with first-principles result, as shown in Fig. 4(c). Moreover, the model can further explain the smaller critical doping density of holes vs electrons in monolayer α-RuCl₃. As shown in Fig. 4 (d), the product of PDOS $\rho_d$ and $\rho_p$ increases more rapidly by hole doping, giving rise to the sharper change of hole doping curve and a smaller critical doping density in Fig. 4(c).

It should be pointed out that the necessary condition for our model is essentially that the energy difference between the FM and AFM orders is small or the existence of energetically competition magnetic orders in candidate materials. In real 2D vdW materials, because of the weak interlayer coupling, this doping effect is easier to affect the interlayer magnetic order. Moreover, our proposed effect is significant in 2D structures because of their enhanced van Hove singularities at band edges and DOS, as seen from Eq. (2).

## V. Doping Driven FM to AFM Transition

The above calculations and analysis show that the FM coupling is always favored by carrier doping. Previous DFT calculations predicted that excessive V atoms at monolayer VI$_3$ will lead to structure transition from VI$_3$ to VI$_2$, and turn it from FM ground states to AFM [60]. It is interesting to ask whether a light/moderate carrier doping can favor the AFM coupling without changing the basic structures of materials. Based on the mechanism in Eq. (2), this idea is possible if more than one type of magnetic ions are involved.

The superexchange paths shown in Fig. 1(a) require two partially occupied $d$ orbitals with the same spin. On the other hand, below these partially occupied states, there could have fully occupied states which may have an opposite direction of magnetic moments from different types of $d$ ions, as shown in Figs. 5(a) and 5(b) for the example of electron doping. For one type of $d$ ions (left part of Figs. 5 (a) and 5(b)), the $t_{2g}$ and $e_g$ orbitals are occupied by less than five electrons, and the partially occupied $d$ orbital has the same spin with its $t_{2g}$ and $e_g$ due to the Hund's rules. For the other type of $d$ ions (right part of Figs. 5 (a) and 5 (b)), the $t_{2g}$ and $e_g$ orbitals are occupied by more than five electrons, and the partially occupied $d$ orbitals must be $t_{2g}^*$ or $e_g^*$, which has the opposite spins between $t_{2g}$ and $e_g$ orbitals due to the Pauli exclusion principle. Because the superexchange path requires two partial occupied $d$ orbitals with the same spins, the AFM coupling (Figs. 5(a)) is preferred while the FM coupling (Fig. 5(b)) is forbidden. It should be mentioned that this match of $d$ orbitals is necessary because doped carriers need to the change occupations of $d$ orbitals of both two different magnetic ions.

Because interlayer coupling is weak, 2D magnetic heterostructures are better candidates than monolayer systems to achieve this FM-AFM transition. We proposed a vdW heterostructure based on the MBT family of materials, TiBi$_2$Te$_4$/NiBi$_2$Te$_4$, to demonstrate this mechanism by DFT+U simulations. Their atomic structures are presented in Fig. 5(c). Intrinsically, there are two electrons occupying the $t_{2g}$ and $e_g$ orbitals of the Ti$^{2+}$ ion. While there are eight electrons occupied the $d$ orbitals of the Ni$^{2+}$ ions, in which there are totally five occupied $t_{2g}$ and $e_g$ orbitals. This means the TiBi$_2$Te$_4$/NiBi$_2$Te$_4$ vdW heterostructure satisfies the conditions stated in the above paragraph, and electron doping may favor AFM interaction if the doped $d$ orbitals of Ti$^{2+}$ and Ni$^{2+}$ have overlaps.

The calculated PDOS of TiBi$_2$Te$_4$/NiBi$_2$Te$_4$ are shown in Fig. 5(d), from which we can see that there are overlapping of PDOS between $d$ orbitals from Ti$^{2+}$ and Ni$^{2+}$ ions at the fermi energy without doping. Importantly, under electron doping, the product of these two PDOSs increases near the fermi energy, as shown by the dashed curves in Fig. 5(d). This indicates that electron doping may favor the interlayer AFM interaction in this system. The DFT-calculated total energy difference between FM and AFM magnetic couplings according to carrier doping is shown in Fig. 5(e). The ground state slightly prefers the interlayer FM without doping, which is about 0.25 meV/u.c, lower than that of AFM. After introducing extra electrons, the doped system favors the interlayer AFM coupling and can eventually tune the magnetic ground state from FM to AFM after a critical doping density of 0.02 e/u.c. (~$1.2 \times 10^{13}$ cm$^{-2}$).

We can quantitatively understand the favorite energy of the AFM state by carrier doping according to Eq. (1). For the TiBi$_2$Te$_4$/NiBi$_2$Te$_4$ heterostructure, Eq. (1) should be written as

$$\Delta J_S \propto t_{dp}^2 [\rho_{Ti(d)}]^{1/2} [\rho_{Ni(d)}]^{1/2}, \qquad (3)$$

where the first part represents the contribution of $d$ orbitals from TiBi$_2$Te$_4$, and the second part from that of NiBi$_2$Te$_4$. The PDOS of $p$ orbitals from Bi and Te ions are small near the Fermi energy, which means the strength of $J_S$ should mainly be contributed by $\rho_{Ti(d)}$ and $\rho_{Ni(d)}$. A fitting formula of the energy difference is $\Delta E = \left\{ t_{dp}^2 \left[ (\rho_{Ti(d)} \rho_{Ni(d)})^{\frac{1}{2}} - (\rho_{Ti(d)}^0 \rho_{Ni(d)}^0)^{\frac{1}{2}} \right] + \Delta E_0 \right\}$, in the unit of meV, which is plotted in in Fig. 5(e). In this formula, $\rho_{Ti(d)}^0$ and $\rho_{Ni(d)}^0$ are the weights of PDOS of $d$ orbitals with $Ti^{2+}$ and $Ni^{2+}$ ions, respectively, at the Fermi energy without carrier doping. The last term ($\Delta E_0$) is used to fit the intrinsic energy difference between AFM and FM states without doping, which is 0.25 meV for $TiBi_2Te_4$/$NiBi_2Te_4$ heterostructure. The hopping factor $t_{dp}^2$ is set to be -2 eV$^2$. The minus sign indicates that this superexchange interaction favors the AFM coupling. As shown in Fig. 5(e), the model excellently matches the DFT-calculated results, especially for electron doping. This verifies our prediction that carrier doping can also favorite the AFM coupling via involving more than one type of magnetic ions.

Finally, it is worth mentioning that the first-principles DFT+U approach may be questionable to satisfactorily describe correlation effects in general 2D magnetic structures. However, it just provides PDOS and wavefunction overlaps for the mechanism described in Figs. 5 (a) and (b). More advanced approaches, such as the GW approximation [61-63] and dynamical Mean-field theory [64], can provide improve inputs.

## VI. CONCLUSIONS

In conclusion, based on the picture of spin superexchange, we have proposed a model based on superexchange interaction to understand the role of doped carriers in tuning the magnetic coupling in 2D magnetic materials. We showed that there are three types of tuning the magnetic exchange couplings in 2D magnetic semiconductors materials by carrier doping, depended on the distribution of correlated $d$ orbitals around the band edges. Furthermore, by using first-principles DFT calculations, we show that this model quantitatively agrees with the DFT-calculated changes of magnetic states in various 2D magnetic materials. Finally, we predict that carrier doping can drive the FM to AFM magnetic transition by including more than one type of magnetic ions.

**ACKNOWLEDGMENTS**

Y.L. is supported by the National Natural Science Foundation of China (grant No. 12164026). H.W. and L.Y. are supported by the National Science Foundation (NSF) grant No. DMR-2124934. L.W. acknowledges the support from Jiangxi Provincial Innovation Talents of Science and Technology under grant No. 20165BCB18003. The computational resources for H.W. and L.Y. are provided by the Extreme Science and Engineering Discovery Environment (XSEDE), which is supported by National Science Foundation (NSF) Grant No. ACI-1548562. H.W. and L.Y. acknowledge the Texas Advanced Computing Center (TACC) at The University of Texas at Austin for providing HPC resources.

**References**

[1] F. Matsukura, Y. Tokura, and H. Ohno, Nat Nanotechnol **10**, 209 (2015).
[2] K. S. Novoselov, A. Mishchenko, A. Carvalho, and A. H. Castro Neto, Science **353**, aac9439 (2016).
[3] S. Jiang, L. Li, Z. Wang, K. F. Mak, and J. Shan, Nat Nanotechnol **13**, 549 (2018).
[4] I. A. Verzhbitskiy, H. Kurebayashi, H. Cheng, J. Zhou, S. Khan, Y. P. Feng, and G. Eda, Nature Electronics **3**, 460 (2020).
[5] Y. Deng, Y. Yu, Y. Song, J. Zhang, N. Z. Wang, Z. Sun, Y. Yi, Y. Z. Wu, S. Wu, J. Zhu, J. Wang, X. H. Chen, and Y. Zhang, Nature **563**, 94 (2018).
[6] Z. Wang, T. Zhang, M. Ding, B. Dong, Y. Li, M. Chen, X. Li, J. Huang, H. Wang, X. Zhao, Y. Li, D. Li, C. Jia, L. Sun, H. Guo, Y. Ye, D. Sun, Y. Chen, T. Yang, J. Zhang, S. Ono, Z. Han, and Z. Zhang, Nat Nanotechnol **13**, 554 (2018).
[7] S. Y. Park, D. S. Kim, Y. Liu, J. Hwang, Y. Kim, W. Kim, J. Y. Kim, C. Petrovic, C. Hwang, S. K. Mo, H. J. Kim, B. C. Min, H. C. Koo, J. Chang, C. Jang, J. W. Choi, and H. Ryu, Nano Lett **20**, 95 (2020).
[8] B. Huang, G. Clark, D. R. Klein, D. MacNeill, E. Navarro-Moratalla, K. L. Seyler, N. Wilson, M. A. McGuire, D. H. Cobden, D. Xiao, W. Yao, P. Jarillo-Herrero, and X. Xu, Nat Nanotechnol **13**, 544 (2018).
[9] G. Zheng, W. Q. Xie, S. Albarakati, M. Algarni, C. Tan, Y. Wang, J. Peng, J. Partridge, L. Farrar, J. Yi, Y. Xiong, M. Tian, Y. J. Zhao, and L. Wang, Phys Rev Lett **125**, 047202 (2020).
[10] Y. L. Han, S. Y. Sun, S. F. Qi, X. H. Xu, and Z. H. Qiao, Physical Review B **103**, 245403 (2021).
[11] Y. Tian, W. Gao, E. A. Henriksen, J. R. Chelikowsky, and L. Yang, Nano Lett **19**, 7673 (2019).
[12] Y. Kasahara, T. Ohnishi, Y. Mizukami, O. Tanaka, S. Ma, K. Sugii, N. Kurita, H. Tanaka, J. Nasu, Y. Motome, T. Shibauchi, and Y. Matsuda, Nature **559**, 227 (2018).
[13] T. Yokoi, S. Ma, Y. Kasahara, S. Kasahara, T. Shibauchi, N. Kurita, H. Tanaka, J. Nasu, Y. Motome, C.

Hickey, S. Trebst, and Y. Matsuda, Science **373**, 568 (2021).

[14] J. A. N. Bruin, R. R. Claus, Y. Matsumoto, N. Kurita, H. Tanaka, and H. Takagi, Nature Physics **18**, 401 (2022).

[15] M. Yamashita, J. Gouchi, Y. Uwatoko, N. Kurita, and H. Tanaka, Physical Review B **102**, 220404 (2020).

[16] H. Li, H. K. Zhang, J. Wang, H. Q. Wu, Y. Gao, D. W. Qu, Z. X. Liu, S. S. Gong, and W. Li, Nat Commun **12**, 4007 (2021).

[17] H. Suzuki, H. Liu, J. Bertinshaw, K. Ueda, H. Kim, S. Laha, D. Weber, Z. Yang, L. Wang, H. Takahashi, K. Fursich, M. Minola, B. V. Lotsch, B. J. Kim, H. Yavas, M. Daghofer, J. Chaloupka, G. Khaliullin, H. Gretarsson, and B. Keimer, Nat Commun **12**, 4512 (2021).

[18] H. Wang, F. Fan, S. Zhu, and H. Wu, EPL (Europhysics Letters) **114**, 47001 (2016).

[19] C. Wang, X. Zhou, Y. Pan, J. Qiao, X. Kong, C.-C. Kaun, and W. Ji, Physical Review B **97**, 245409 (2018).

[20] G. Kresse and J. Furthmuller, Physical Review B **54**, 11169 (1996).

[21] G. Kresse and D. Joubert, Physical Review B **59**, 1758 (1999).

[22] J. P. Perdew, A. Ruzsinszky, G. I. Csonka, O. A. Vydrov, G. E. Scuseria, L. A. Constantin, X. Zhou, and K. Burke, Phys Rev Lett **100**, 136406 (2008).

[23] A. I. Liechtenstein, V. V. Anisimov, and J. Zaanen, Phys Rev B Condens Matter **52**, R5467 (1995).

[24] Y. Lu, R. Fei, X. Lu, L. Zhu, L. Wang, and L. Yang, ACS Appl Mater Interfaces **12**, 6243 (2020).

[25] J. P. Perdew, K. Burke, and M. Ernzerhof, Phys Rev Lett **77**, 3865 (1996).

[26] S. L. Dudarev, G. A. Botton, S. Y. Savrasov, C. J. Humphreys, and A. P. Sutton, Physical Review B **57**, 1505 (1998).

[27] M. M. Otrokov, I. P. Rusinov, M. Blanco-Rey, M. Hoffmann, A. Y. Vyazovskaya, S. V. Eremeev, A. Ernst, P. M. Echenique, A. Arnau, and E. V. Chulkov, Phys Rev Lett **122**, 107202 (2019).

[28] J. Li, Y. Li, S. Du, Z. Wang, B. L. Gu, S. C. Zhang, K. He, W. Duan, and Y. Xu, Sci Adv **5**, eaaw5685 (2019).

[29] S. Grimme, J. Antony, S. Ehrlich, and H. Krieg, J Chem Phys **132**, 154104 (2010).

[30] S. Blundell, *Magnetism in Condensed Matter* (Oxford University Press, Oxford, 2001).

[31] P. W. Anderson, Physical Review **79**, 350 (1950).

[32] J. B. Goodenough, Physical Review **100**, 564 (1955).

[33] J. Kanamori, Journal of Physics and Chemistry of Solids **10**, 87 (1959).

[34] B. Huang, G. Clark, E. Navarro-Moratalla, D. R. Klein, R. Cheng, K. L. Seyler, D. Zhong, E. Schmidgall, M. A. McGuire, D. H. Cobden, W. Yao, D. Xiao, P. Jarillo-Herrero, and X. Xu, Nature **546**, 270 (2017).

[35] T. Song, X. Cai, M. W. Tu, X. Zhang, B. Huang, N. P. Wilson, K. L. Seyler, L. Zhu, T. Taniguchi, K. Watanabe, M. A. McGuire, D. H. Cobden, D. Xiao, W. Yao, and X. Xu, Science **360**, 1214 (2018).

[36] D. R. Klein, D. MacNeill, J. L. Lado, D. Soriano, E. Navarro-Moratalla, K. Watanabe, T. Taniguchi, S. Manni, P. Canfield, J. Fernandez-Rossier, and P. Jarillo-Herrero, Science **360**, 1218 (2018).

[37] Z. Wang, I. Gutierrez-Lezama, N. Ubrig, M. Kroner, M. Gibertini, T. Taniguchi, K. Watanabe, A. Imamoglu, E. Giannini, and A. F. Morpurgo, Nat Commun **9**, 2516 (2018).

[38] B. Wu, J. Yang, R. Quhe, S. Liu, C. Yang, Q. Li, J. Ma, Y. Peng, S. Fang, J. Shi, J. Yang, J. Lu, and H. Du, Physical Review Applied **17**, 034030 (2022).

[39] S. Jiang, J. Shan, and K. F. Mak, Nat Mater **17**, 406 (2018).

[40] T. Song, Z. Fei, M. Yankowitz, Z. Lin, Q. Jiang, K. Hwangbo, Q. Zhang, B. Sun, T. Taniguchi, K. Watanabe, M. A. McGuire, D. Graf, T. Cao, J. H. Chu, D. H. Cobden, C. R. Dean, D. Xiao, and X. Xu, Nat Mater **18**, 1298 (2019).

[41] T. Li, S. Jiang, N. Sivadas, Z. Wang, Y. Xu, D. Weber, J. E. Goldberger, K. Watanabe, T. Taniguchi, C. J. Fennie, K. Fai Mak, and J. Shan, Nat Mater **18**, 1303 (2019).
[42] Z. Sun, Y. Yi, T. Song, G. Clark, B. Huang, Y. Shan, S. Wu, D. Huang, C. Gao, Z. Chen, M. McGuire, T. Cao, D. Xiao, W. T. Liu, W. Yao, X. Xu, and S. Wu, Nature **572**, 497 (2019).
[43] L. Thiel, Z. Wang, M. A. Tschudin, D. Rohner, I. Gutierrez-Lezama, N. Ubrig, M. Gibertini, E. Giannini, A. F. Morpurgo, and P. Maletinsky, Science **364**, 973 (2019).
[44] W. Chen, Z. Sun, Z. Wang, L. Gu, X. Xu, S. Wu, and C. Gao, Science **366**, 983 (2019).
[45] N. Sivadas, S. Okamoto, X. Xu, C. J. Fennie, and D. Xiao, Nano Lett **18**, 7658 (2018).
[46] P. Jiang, C. Wang, D. Chen, Z. Zhong, Z. Yuan, Z.-Y. Lu, and W. Ji, Physical Review B **99**, 144401 (2019).
[47] S. W. Jang, M. Y. Jeong, H. Yoon, S. Ryee, and M. J. Han, Physical Review Materials **3**, 031001 (2019).
[48] D. Soriano, C. Cardoso, and J. Fernández-Rossier, Solid State Communications **299**, 113662 (2019).
[49] D. Zhang, M. Shi, T. Zhu, D. Xing, H. Zhang, and J. Wang, Phys Rev Lett **122**, 206401 (2019).
[50] Y.-J. Hao, P. Liu, Y. Feng, X.-M. Ma, E. F. Schwier, M. Arita, S. Kumar, C. Hu, R. e. Lu, M. Zeng, Y. Wang, Z. Hao, H.-Y. Sun, K. Zhang, J. Mei, N. Ni, L. Wu, K. Shimada, C. Chen, Q. Liu, and C. Liu, Physical Review X **9**, 041038 (2019).
[51] Y. Deng, Y. Yu, M. Z. Shi, Z. Guo, Z. Xu, J. Wang, X. H. Chen, and Y. Zhang, Science **367**, 895 (2020).
[52] H. Sun, B. Xia, Z. Chen, Y. Zhang, P. Liu, Q. Yao, H. Tang, Y. Zhao, H. Xu, and Q. Liu, Phys Rev Lett **123**, 096401 (2019).
[53] C. Hu, L. Ding, K. N. Gordon, B. Ghosh, H. J. Tien, H. Li, A. G. Linn, S. W. Lien, C. Y. Huang, S. Mackey, J. Liu, P. V. S. Reddy, B. Singh, A. Agarwal, A. Bansil, M. Song, D. Li, S. Y. Xu, H. Lin, H. Cao, T. R. Chang, D. Dessau, and N. Ni, Sci Adv **6**, eaba4275 (2020).
[54] Y. Zhao and Q. Liu, Applied Physics Letters **119**, 060502 (2021).
[55] J. Shao, Y. Liu, M. Zeng, J. Li, X. Wu, X. M. Ma, F. Jin, R. Lu, Y. Sun, M. Gu, K. Wang, W. Wu, L. Wu, C. Liu, Q. Liu, and Y. Zhao, Nano Lett **21**, 5874 (2021).
[56] R. Gao, G. Qin, S. Qi, Z. Qiao, and W. Ren, Physical Review Materials **5**, 114201 (2021).
[57] R. D. Johnson, S. C. Williams, A. A. Haghighirad, J. Singleton, V. Zapf, P. Manuel, I. I. Mazin, Y. Li, H. O. Jeschke, R. Valentí, and R. Coldea, Physical Review B **92**, 235119 (2015).
[58] H.-S. Kim and H.-Y. Kee, Physical Review B **93**, 155143 (2016).
[59] F. Iyikanat, M. Yagmurcukardes, R. T. Senger, and H. Sahin, Journal of Materials Chemistry C **6**, 2019 (2018).
[60] M. Baskurt, I. Eren, M. Yagmurcukardes, and H. Sahin, Applied Surface Science **508**, 144937 (2020).
[61] M. Wu, Z. Li, T. Cao, and S. G. Louie, Nat Commun **10**, 2371 (2019).
[62] L. Zhu and L. Yang, Physical Review B **101**, 245401 (2020).
[63] M. Wu, Z. Li, and S. G. Louie, Physical Review Materials **6**, 014008 (2022).
[64] G. Kotliar, S. Y. Savrasov, K. Haule, V. S. Oudovenko, O. Parcollet, and C. A. Marianetti, Reviews of Modern Physics **78**, 865 (2006).

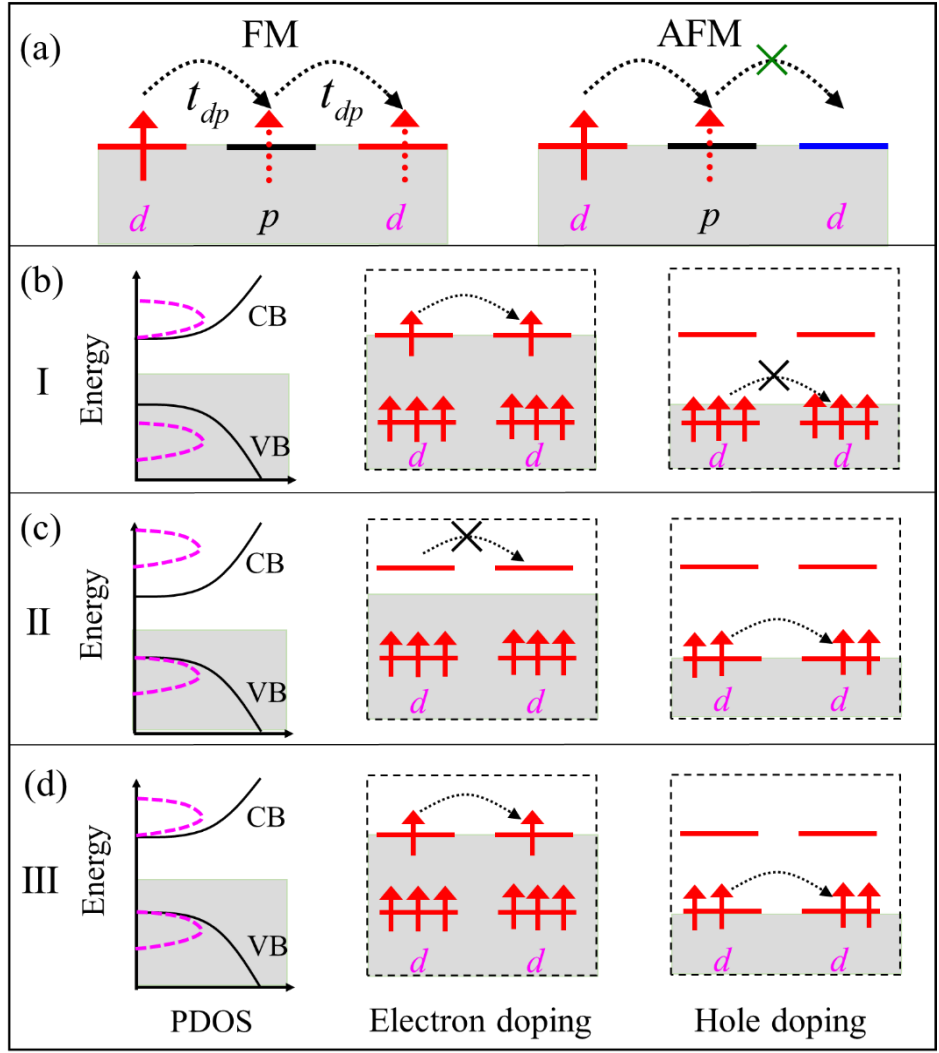

FIG. 1. (a) Spin superexchange path between partially occupied $d$ orbitals at FM coupling. Because of the requirement of spin conservation, the spin AFM superexchange path between partially occupied $d$ orbitals is forbidden, as denoted by the green "×" symbol. Red and blue colors denote spin up and down, respectively. (b) Type I DOS, in which the black-solid parabolic lines and magenta-dashed peaks represent the total band structure and the PDOS of $d$ orbitals, respectively. The fermi level is denoted by the upper boundary of the shadow region. Electron doping introduces partial $d$ orbitals at the conduction edge. However, the valence band edge is mostly from the $p$ orbitals, and hole doping cannot change the $d$-orbital occupation. Thus, spin superexchanges between local $d$ valence orbitals is forbidden due to Fermi blocking under hole doping. (c) Type II DOS, in which electron doping cannot introduce partial $d$ orbitals at the conduction edge, but hole doping can at the valence edge. (d) Type III DOS, in which both electron and hole doping can introduce partial $d$ orbitals at conduction and valence edges, respectively, resulting allowed hopping.

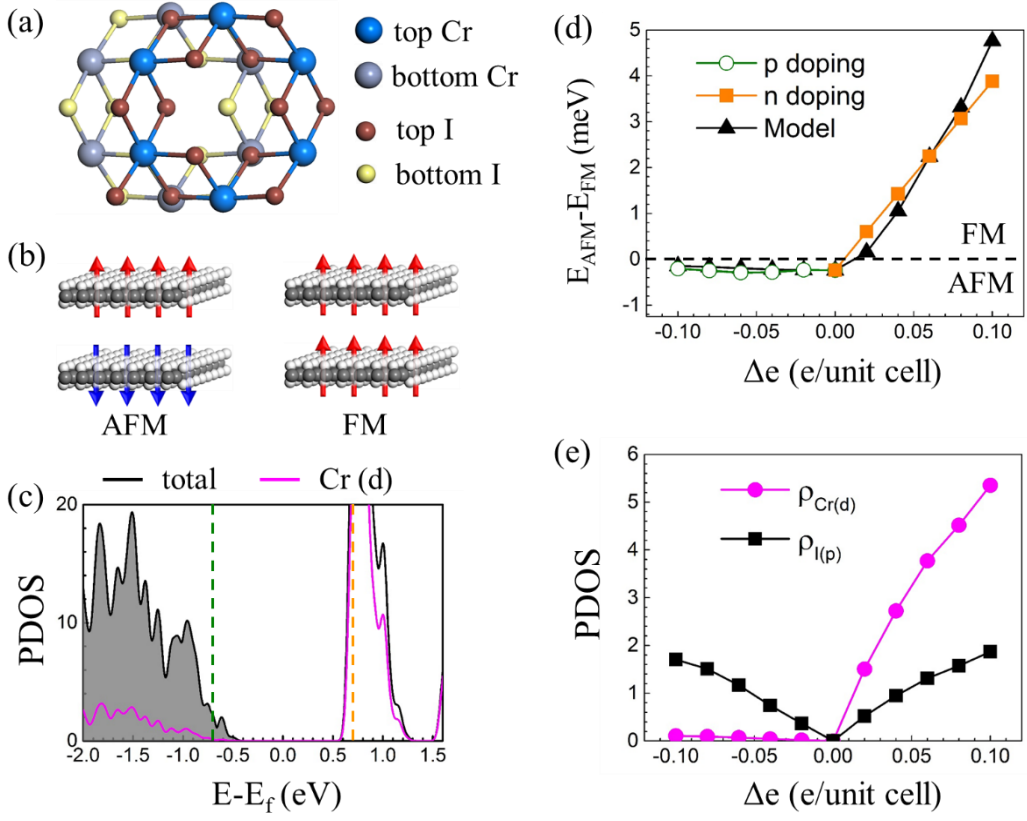

FIG. 2. (a) Structures of bilayer CrI3 at high temperature phase. (b) The two most energy stable magnetic configuration of bilayer CrI$_3$. (c) Distribution of *d* orbitals and total DOS of bilayer CrI$_3$, which imply this DOS belongs to type I. (d) Change of ground state of bilayer CrI$_3$ under carrier doping. Doped electrons can change the magnetic ground state from interlayer AFM to FM coupling, but doped holes cannot. The black triangle curve is fitting according to Eq. (2). (e) Change of heights of PDOS at Fermi energy for *d* orbitals of Cr and *p* orbitals of I under carrier doping.

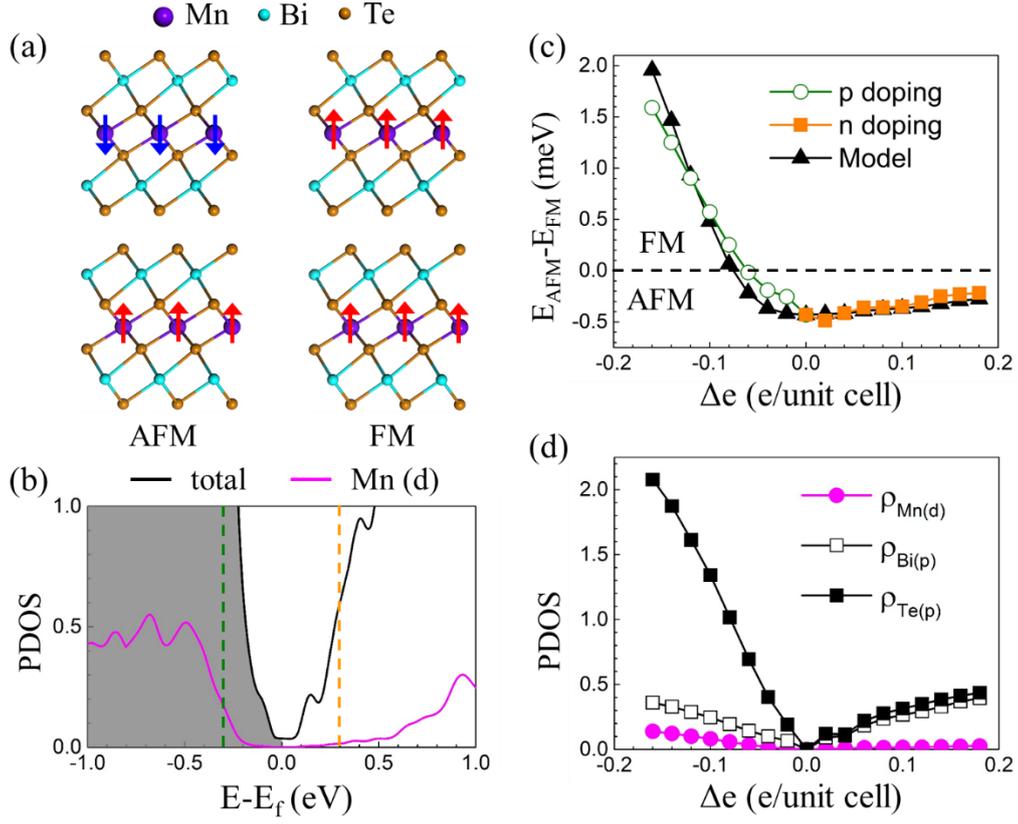

FIG. 3. (a) Structures and the two most energy stable magnetic configuration of bilayer $MnBi_2Te_4$. (b) Distribution of $d$ orbitals and total DOS of bilayer $MnBi_2Te_4$, which imply this DOS belongs to type II. (c) Change of ground state of bilayer $MnBi_2Te_4$ under carrier doping. Doped holes can change the magnetic ground state from interlayer AFM to FM coupling, but doped electrons cannot. The black triangle curve is fitting according to Eq. (2). (d) Change of heights of PDOS at Fermi energy for $d$ orbitals of Mn, $p$ orbitals of Bi, and $p$ orbitals of Te under carrier doping.

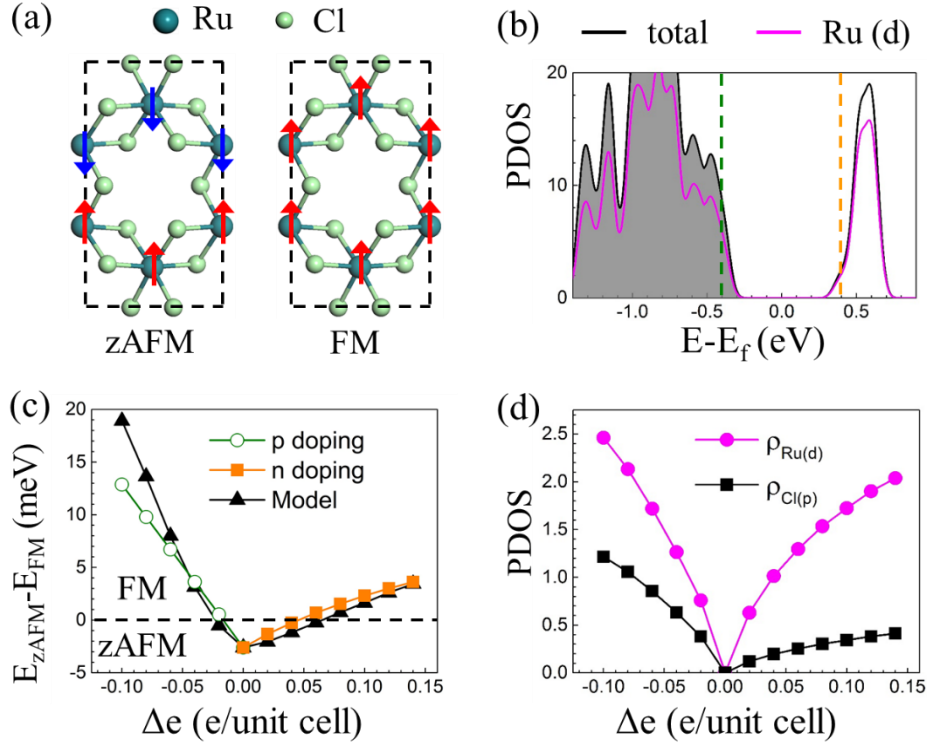

FIG. 4. (a) Structures and the two most energy stable magnetic configuration of monolayer α-RuCl$_3$. (b) Distribution of $d$ orbitals and total DOS of monolayer α-RuCl$_3$, which imply this DOS belongs to type III. (c) Change of magnetic ground state of monolayer α-RuCl3 under carrier doping. Both doped electrons and holes can change the magnetic ground states from zAFM to FM. The black triangle curve is fitting according to Eq. (2). (d) Change of heights of PDOS at Fermi energy for $d$ orbitals of Ru and $p$ orbitals of Cl under carrier doping.

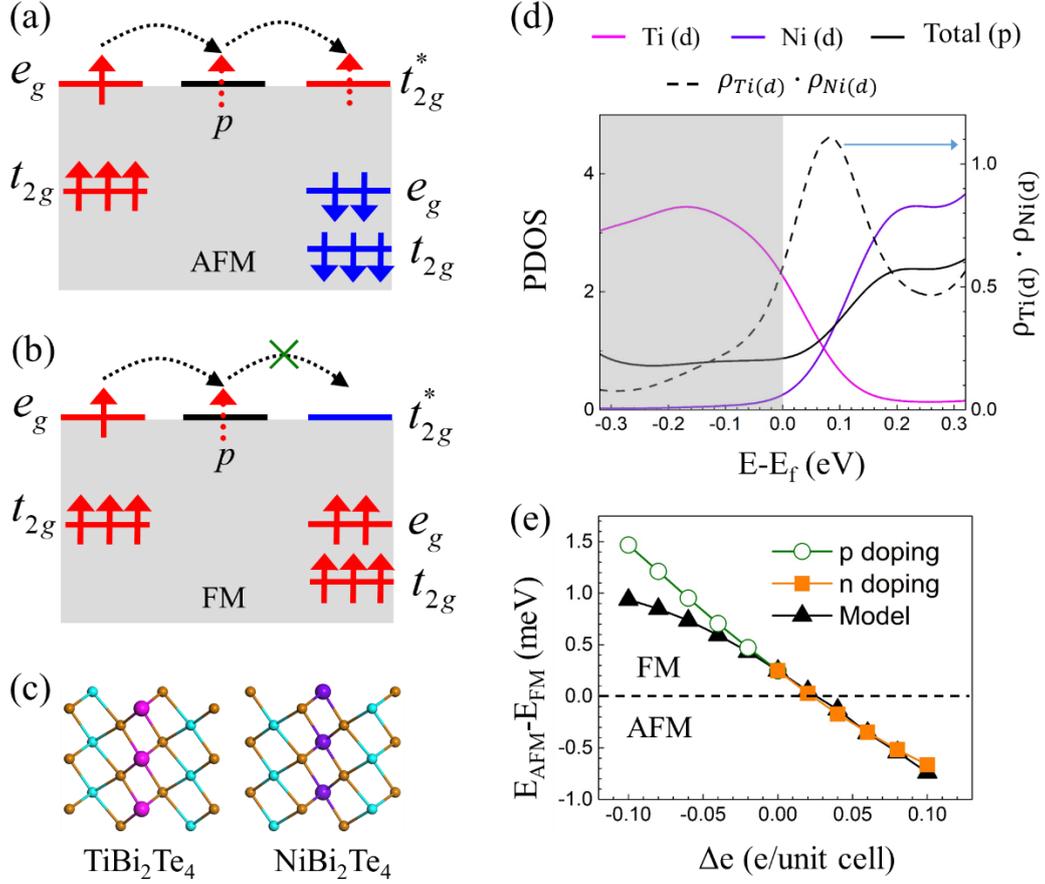

FIG. 5. (a) Spin superexchange path between two different $d$ ions in AFM coupling. (b) Spin superexchange path between two different $d$ ions in FM coupling. (c) The vdW heterostructure formed by $TiBi_2Te_4$ and $NiBi_2Te_4$. (d) PDOS of $TiBi_2Te_4$/$NiBi_2Te_4$ heterostructure. (e) Change of magnetic ground state of this heterostructure under carrier doping. Doped electrons can favorite interlayer AFM coupling in this heterostructure.